\documentclass[a4paper,12pt]{article}
\pdfoutput=1 
\usepackage[fleqn]{amsmath}
\usepackage{jheppub}                                         
\usepackage{color}
\usepackage{rotating}
\usepackage{booktabs}
\usepackage{hvfloat}
\usepackage{float}
\usepackage{adjustbox}
\usepackage{graphicx}
\usepackage[left=4.5cm,right=1.5cm,top=5.5cm]{geometry}

\newcommand{\be}{\begin{equation}}
\newcommand{\ee}{\end{equation}}
\newcommand{\ba}{\begin{eqnarray}}
\newcommand{\ea}{\end{eqnarray}}

\newcommand{\beq}{\begin{equation}}
\newcommand{\eeq}{\end{equation}}
\newcommand{\beqa}{\begin{eqnarray}}
\newcommand{\eeqa}{\end{eqnarray}}
\newcommand{\nn}{\nonumber}

\title{Reverse Hawking-Page Phase Transition in de Sitter Black Holes}

\author{Saoussen Mbarek and Robert B. Mann}
\emailAdd{smbarek@uwaterloo.ca}
\emailAdd{rbmann@uwaterloo.ca}
\affiliation{Department of Physics and Astronomy, University of Waterloo,
Waterloo, Ontario, Canada, N2L 3G1}

\abstract{
In the context of black hole chemistry, we study the thermodynamics of asymptotically de Sitter black holes with conformal scalar hair in Einstein gravity. The hair parameter allows us to attain thermodynamic equilibrium between the event horizon and the cosmological horizon. We find that the system of the black hole and the de Sitter space surrounding it undergo a ``\textit{Reverse}'' Hawking-Page phase transition provided we consider the grand-canonical ensemble.}

\begin{document} 
\maketitle
\flushbottom


\section{Introduction}
The study of Black Holes as thermodynamic systems has been an increasingly active  area of research in the last couple of decades. Investigations of asymptotically anti de-Sitter (AdS) black holes have been at the forefront in advancing our understanding of this subject in large part because  their thermodynamic behaviour has been found to be  analogous to everyday life thermodynamic systems, a subject known as Black Hole Chemistry \cite{Kubiznak:2014zwa}.

In Black Hole Chemistry the cosmological constant $\Lambda$ is treated as a thermodynamic variable extending the phase space of black hole thermodynamics  \cite{CreightonMann:1995}.  The mass of the black hole can be understood as enthalpy  \cite{Kastor2009} and the cosmological constant as pressure, with a conjugate thermodynamic volume $V$ \cite{Caldarelli2000, Kastor2009, Dolan2010, Dolan2011a, Dolan2011, Dolan2012, Cvetic2010, Larranaga2011, Larranaga2012,Gibbons2012, Kubiznak2012, Gunasekaran2012, Belhaj2012,  Lu2012, Smailagic2012, Hendi2012}.   From this viewpoint it has been shown that the Hawking-Page phase transition \cite{Hawking:1982dh}  is analogous to a solid/liquid phase transition  \cite{Kubiznak:2014zwa}, and  that the 4-dimensional Reissner-Nordstr\"om AdS black hole can be interpreted as a Van der Waals (VdW) fluid with the same critical exponents \cite{Kubiznak:2012wp}.    Along with  more general VdW behaviour with standard critical exponents  \cite{Dolan2013}, a broad range of other `chemical' black hole behaviour was subsequently discovered, including  reentrant phase transitions \cite{Altamirano:2013ane}, tricritical points \cite{Altamirano:2013uqa}, Carnot cycles \cite{Johnson2014a}, isolated critical points \cite{Frassino2014,Dolan:2014vba}, extensions to black rings \cite{Altamirano:2014tva}, and superfluidity \cite{Hennigar:2016xwd}.

Our knowledge of the thermodynamic behaviour of asymptotically de Sitter (dS) black holes, for which $\Lambda>0$,  is  significantly more sparse \cite{Romans1992,MannRoss:1995,Li:2014ixn,Tian:2014ila,Azreg-Ainou:2014lua,Kubiznak:2015bya,Zhang:2016yek}. 
Yet their importance to cosmology and to a posited  duality between gravity in de Sitter space and conformal field theory \cite{Strominger:2001pn} make them important objects of investigation.  However this is a complex problem, since the absence of a Killing vector that is everywhere timelike outside the black hole horizon renders   a good notion of the asymptotic mass questionable. Furthermore, the presence of both a black hole horizon and a cosmological horizon yields two distinct temperatures, suggesting that the system is in a non-equilibrium state. 

There have been a few attempts to overcome this issue and study the thermodynamics of black holes in de Sitter spacetime. One approach involves studying the horizons of the de Sitter spacetime separately, considering them as separate thermodynamic systems characterized by their own temperature and thermodynamic behaviours  \cite{Dolan:2013ft, Kubiznak:2015bya, Cai:2001sn, Cai:2001tv, Sekiwa:2006qj}. Others have investigated putting the black hole inside a cavity and imposing  thermodynamic equilibrium in this closed system 
\cite{Carlip2003, Braden:1990hw,Simovic:2018tdy}. This solves the thermodynamic equilibrium problem at the price of  isolating the cosmological horizon and ignoring its contribution to thermodynamics of the system.

Here we present a new approach that overcomes the two-horizon problem in de Sitter spacetime and successfully studies its black hole chemistry. By adding hair to the black hole we find that we can maintain thermodynamic equilibrium  between the two horizons. Other conserved quantities (such as charge) can be added to the black hole
and it becomes possible to explore a range of black hole phase transitions in de Sitter spacetime.

This is the task of this paper.  We find considerably different behaviour for charged hairy de Sitter black holes than we do for their Anti de Sitter (AdS) counterparts \cite{Hennigar:2016xwd,Hennigar:2016ekz,Dykaar:2017mba}. In specific terms, we find that  the system can undergo a phase transition that resembles the Hawking-Page phase transition, but we do not find any swallowtail structure in the free-energy, signatory of a first order phase transition from large to small black holes.  Our results are commensurate with studies of de Sitter black holes in a cavity \cite{Carlip2003, Braden:1990hw}, though  significant details differ once the thermodynamic phase space
is extended to include pressure \cite{Simovic:2018tdy}.

Our paper is organized as follows.  In the next section  we briefly review the basics of conformally coupling scalar fields to gravity and their resultant hairy black holes solutions.  In section 3 we specialize to the case of charged hairy black holes in de Sitter spacetime.  When considering the phase behaviour of these systems, we employ the extended thermodynamic phase space formalism to study how their thermodynamic parameters behave at constant pressure (cosmological constant) and at constant ``chemical'' potential. Furthermore, in a search of possible phase transitions, we study the behaviour of the free energy in different ensembles. We find that a system of a charged hairy black hole in de Sitter will undergo a \textit{Reverse} Hawking-Page phase transition if studies in the grand-canonical ensemble, but won't undergo any phase transitions if studied in the canonical ensemble. The latter is due to a violation of the conservation of charge. 


\section{Conformal Scalar Coupling and Hairy Black Hole Thermodynamics}

Although we are interested in Einstein gravity, we shall set our investigation in the context of  Lovelock gravity minimally coupled to a Maxwell field. The  scalar hair is conformally coupled to gravity via the dimensionally extended Euler densities in terms of the rank four tensor  \citep{Oliva:2011np}
\begin{align}\
{S_{\mu \nu}}^{\gamma \delta} &= \phi^2 {R_{\mu \nu}}^{\gamma \delta} - 2 {\delta}_{[ \mu}^{[ \gamma} {\delta}_{\nu]}^{\delta]} \nabla_{\rho} \phi \nabla^{\rho}  \phi - 4 \phi {\delta}_{[ \mu}^{[ \gamma} \nabla_{\nu]} \nabla^{\delta]} \phi + 8 {\delta}_{[ \mu}^{[ \gamma} \nabla_{\nu]} \phi \nabla^{\delta]} \phi \
\label{4tensor}
\end{align}
where $\phi$ is the scalar field. Under a conformal transformation $g_{\mu \nu} \rightarrow \Omega^2g_{\mu \nu}$  and  $\phi \rightarrow \Omega^{-1} \phi$ the tensor 
${S_{\mu \nu}}^{\gamma \delta} \rightarrow \Omega^4 {S_{\mu \nu}}^{\gamma \delta}$. The action is 
\be \label{action} 
{\cal I} =  \frac{1}{16 \pi G}\int d^dx \sqrt{-g} \, \left( \sum_{k=0}^{k_{\rm max}} {\cal L}^{(k)} - 4 \pi G F_{\mu\nu}F^{\mu\nu} \right)
\ee
where 
\begin{align}
{\cal  L}^{(k)} = &  \frac{1}{2^k} \delta^{(k)} \left(a_k \prod_r^k R^{{\alpha_{r}}{\beta_{r}}}_{{\mu_r} {\nu_r}} \right. \left. + b_k \phi^{d-4k}  \prod_r^k S^{{\alpha_{r}}{\beta_{r}}}_{{\mu_r} {\nu_r}}  \right)
\end{align}
with $\delta^{(k)} = \delta^{\alpha_1 \beta_1 \cdots \alpha_k \beta_k}_{\mu_1 \nu_1 \cdots \mu_k \nu_k}$ the generalized Kronecker tensor,  $a_k$ and $b_k$ are coupling constants, and $k_{\rm max} \le (d-1)/2$. 

The corresponding spherically symmetric topological black hole solutions to the metric of this theory are called Hairy Black Holes \cite{Hennigar:2016ekz}. In a $d$ dimensional spacetime, this metric is of the form
\begin{equation}
ds^{2} = -f dt^{2}+f^{-1} dr^{2}+r^{2} d\Sigma _{\sigma \left(d-2\right)}^{2}
\label{General metric}
\end{equation}
where $d\Sigma _{\sigma \left(d-2\right)}^{2}$ is the line element on a hypersurface of constant scalar curvature that corresponds to the flat, spherical and hyperbolic horizon geometries for $\sigma = 0, +1, -1$, respectively. The volume of this submanifold, $\omega _{ \left(d-2\right)}^{(+1)}  =  {2 \pi^{(d-1)/2}}/{\Gamma\left( \frac{d-1}{2} \right)} $, is simply the volume of a sphere for $\sigma = +1$. The field equations of this theory of gravity give a solution provided $f$ solves the following polynomial equation \cite{Hennigar:2016ekz}
\begin{align}\
\sum_{k=0}^{k_{max}} \alpha_{k} \left( \frac{\sigma - f}{r^2} \right)^k &= \frac{16 \pi G M} {(d-2) \omega_{ \left(d-2\right)}^{(\sigma)} r^{d-1}} + \frac{H}{r^d} - \frac{8 \pi G}{(d-2)(d-3)} \frac{Q^2}{r^{2d-4}}\
\label{general_Poly}
\end{align}
where $M$, $H$ and $Q$ are the mass, hair parameter and charge respectively, and
\begin{align} \label{coupling}
\alpha_0 &= \frac{a_0}{(d-1)(d-2)}\qquad  \alpha_1 = a_1 \qquad  \alpha_k = a_k \prod_{n=3}^{2k} (d-n) \,\textrm{ for } k \ge 2\, ,
\end{align}
 For consistency with \cite{Hennigar:2016ekz}, we shall set $\alpha_1 = 1$ and $a_0 = -2 \Lambda < 0$ to recover general relativity at the limit $\alpha_k \rightarrow 0$ for $k > 1$ and we also set $G=1$. 
We also have
\be\label{defn_of_H} 
\phi = \frac{N}{r} \qquad
H = \sum_{k=0}^{k_{\rm max}} \frac{(d-3)!}{(d-2(k+1))!}b_k \sigma^k N^{d-2k} \, 
\ee
the respective scalar field and ``hair parameter''. To satisfy the equations of motion the integration constant $N$ must satisfy the following constraints
\begin{align} \label{constraints}
\sum_{k=1}^{k_{max}} k b_{k} \frac{(d-1)!}{(d-2k-1)!} \sigma^{k-1} N^{2-2k}&= 0  \nn\\
\sum_{k=0}^{k_{max}} b_{k} \frac{(d-1)!(d(d-1)+4k^2)}{(d-2k-1)!} \sigma^{k} N^{ -2k}&= 0 
\end{align}
Since N is  the only unknown in \eqref{constraints}, then one of these equations plays the role of a constraint on the permitted coupling constants .

For asymptotically de Sitter solutions $a_0<0$, and a black hole solution for $f$ from
\eqref{general_Poly} will have at least two horizons: a cosmological horizon at 
$r=r_c$ and a black hole horizon at $r=r_+$.  

To investigate the thermodynamics of these black holes we need to compute their temperature and entropy. For the latter,
as shown in \cite{Hennigar:2016ekz}, we use Wald's method \cite{Wald:1993nt}, obtaining
\be 
S_h = \frac{\Sigma^{\sigma}_{d-2}}{4 G}   \left[ \sum_{k=1}^{k_{\rm max}} \frac{(d-2) k \sigma^{k-1}  \alpha_k   r_h^{d-2k}}{d-2k}  - \frac{d H}{2\sigma (d-4)}\right] 
\label{entropy}
\ee
where $r_h \in \{r_c,r_+\}$ is the horizon size and we have set  $b_k = 0 \, \, \forall \ \ k > 2$ for simplicity. Recalling that temperature $T_h =  \left\vert\frac{f'(r_h)}{4 \pi}\right\vert$, 
it is straightforward to verify that both the (extended) first law of thermodynamics and the Smarr relation hold at both horizons \cite{Dolan:2013ft}.  For the black hole these read respectively
\begin{align}  
 \delta M_+ &= T_+ \delta S_+ + V_+ \delta P_+ + \Phi_+ \delta Q + \kappa \delta H 
\label{black hole_FL_Smarr1} \\
 (d - 3) M_+ &= (d - 2) T_+ S_+ - 2 V_+ P_+ +(d - 3) \Phi_+ Q + (d - 2) \kappa H
 \label{black hole_FL_Smarr2}
\end{align}
where the subscript `$+$' refers to the black hole.  
Similarly, but not identically, the extended first law and the Smarr relation corresponding to the cosmological horizon are respectively
\begin{align} 
\ \delta M_c &= - T_c \delta S_c + V_c \delta P_c + \Phi_c \delta Q + \kappa \delta H 
\label{Cosmo_FL_Smarr1}  \\
 (d - 3) M_c &= - (d - 2) T_c S_c - 2 V_c P_c +(d - 3) \Phi_c Q + (d - 2) \kappa H
 \label{Cosmo_FL_Smarr2} 
\end{align}
where the subscript `$c$' refers to the cosmological horizon.

The key feature we will exploit in examining the thermodynamics of these black holes is that the additional
degree of freedom from the hair parameter allows us to require that the temperatures at both horizons 
be equal  
\begin{equation}
T_+ = \left\vert\frac{f'(r_+)}{4 \pi} \right\vert  =  \left\vert\frac{f'(r_c)}{4 \pi} \right\vert  =T_c = T
\label{5d_Temp}
\end{equation}
equilibrating the particle flux at both horizons. This  
ensures thermodynamic equilibrium whilst retaining the same numbers of thermodynamic degrees of freedom
present in the Reissner-Nordstrom AdS black hole.  In the sequel we shall consider the thermodynamics of these charged hairy black holes in Einstein gravity.


\section{Asymptotically de Sitter Hairy black holes}

With the tools developed in the previous sections, we are now situated to perform the extended phase space analysis for these systems.  As we want to see if there are any hidden phase transitions that either the black hole or the full system undergo, we will not study the thermodynamic behaviour of each and every parameter, but rather focus on analyzing the Gibbs free energy, with the equilibrium state being the global minimum of this quantity.
The latter can be considered in the grand-canonical ensemble, in which  the charge is considered as  variable and the potential is fixed. Alternatively we can consider the canonical ensemble in which charge is the parameter that is fixed. 
\begin{figure}[H]
\centering
\includegraphics[width=0.4\textwidth]{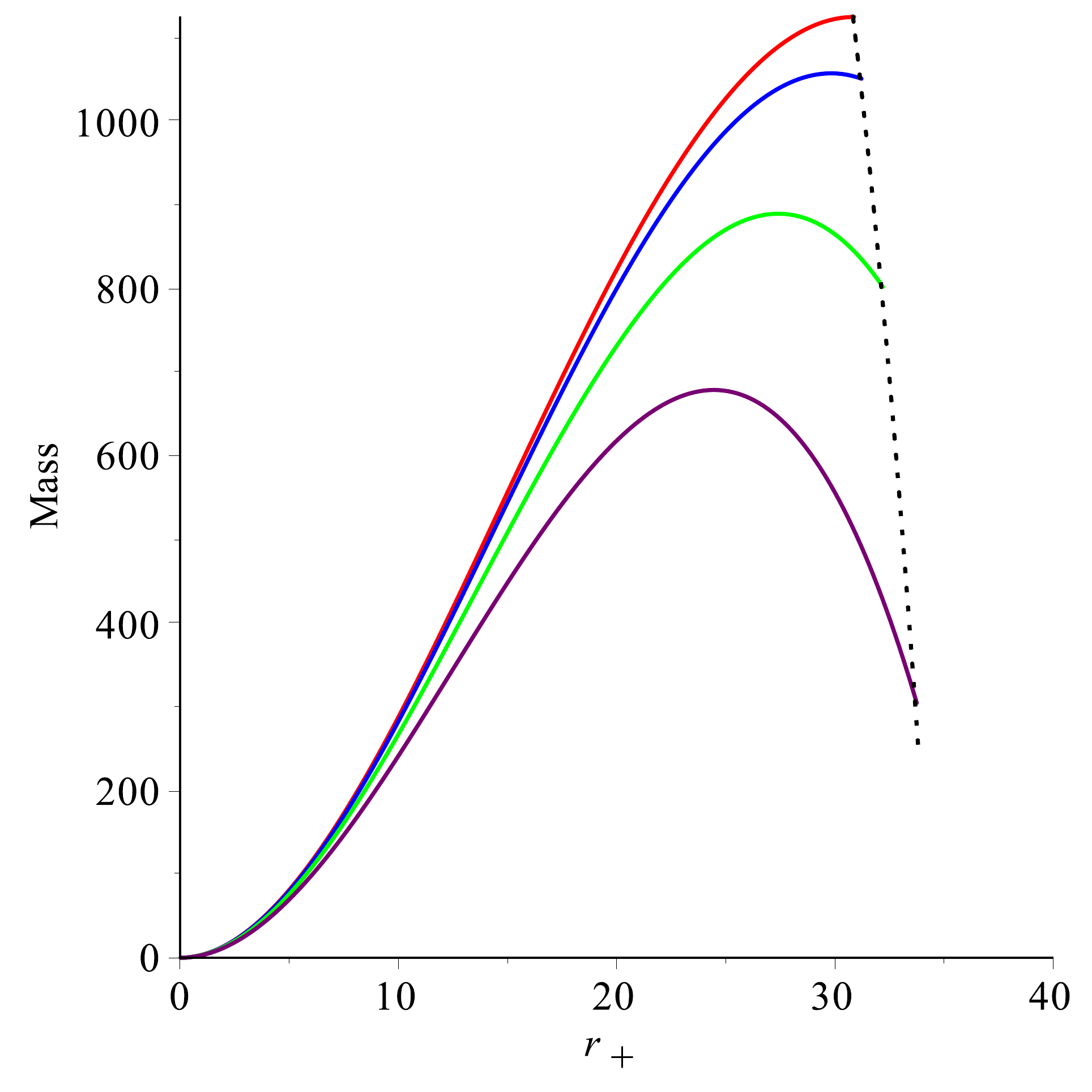}
\includegraphics[width=0.4\textwidth]{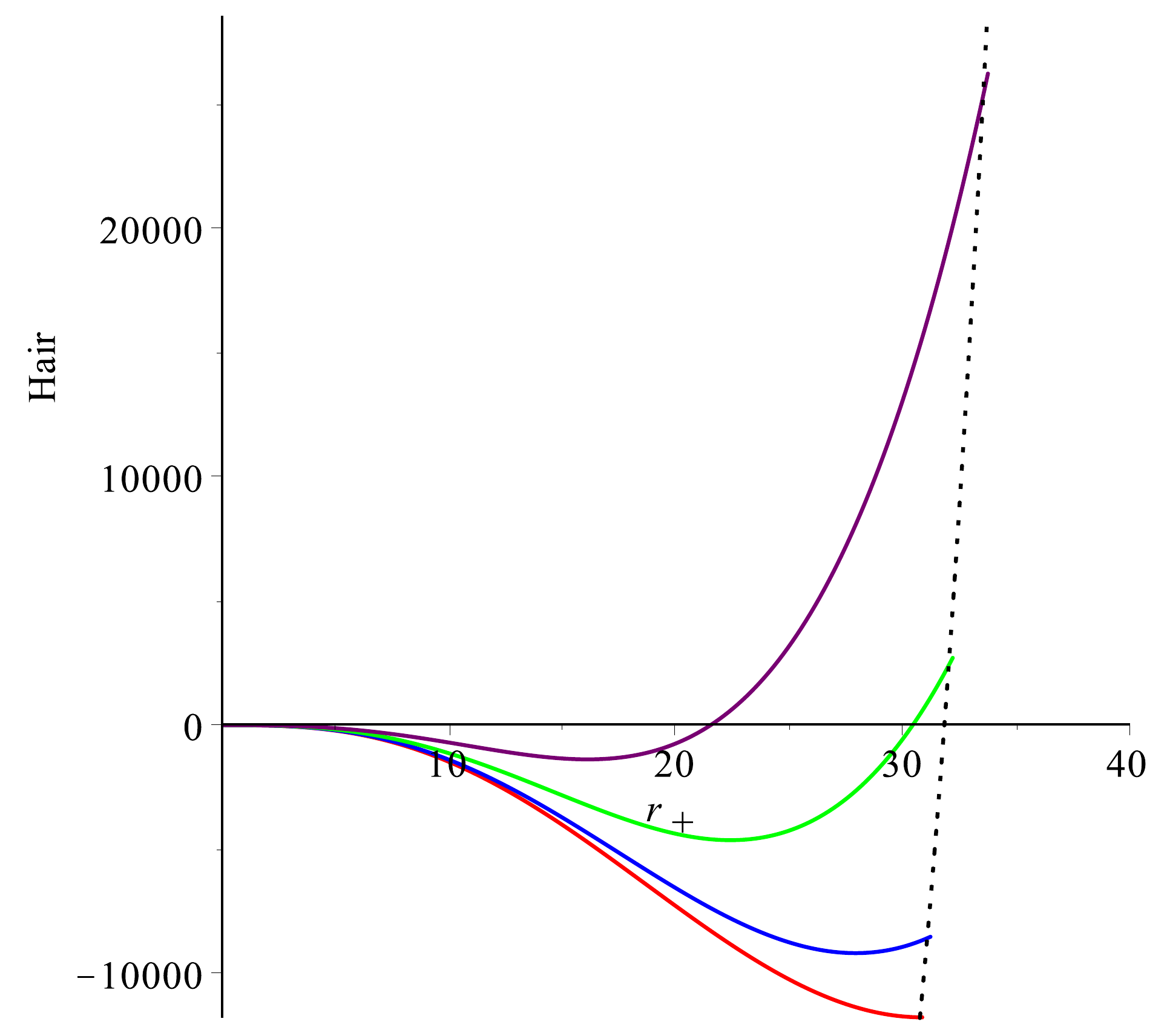}
\includegraphics[width=0.4\textwidth]{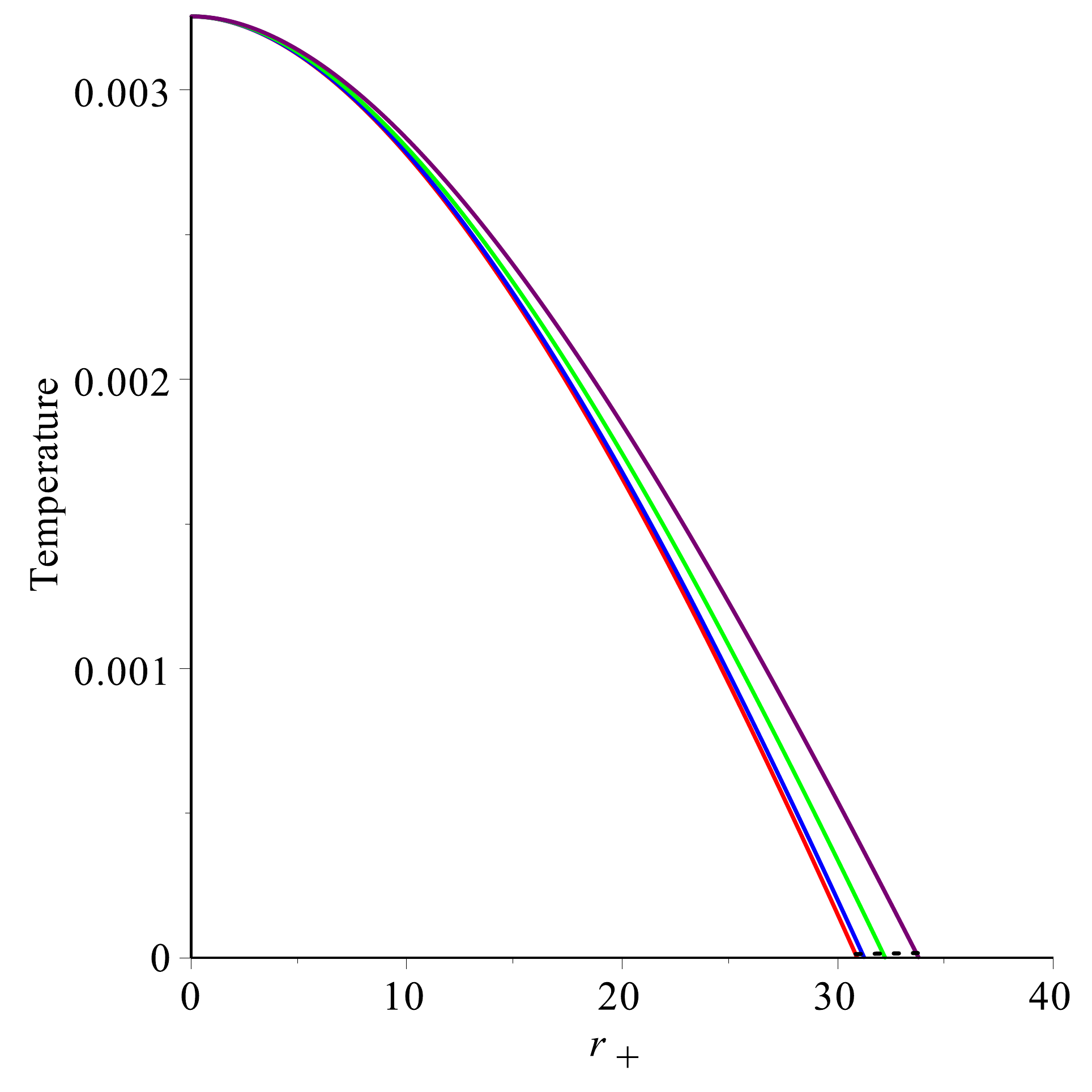}
\includegraphics[width=0.4\textwidth]{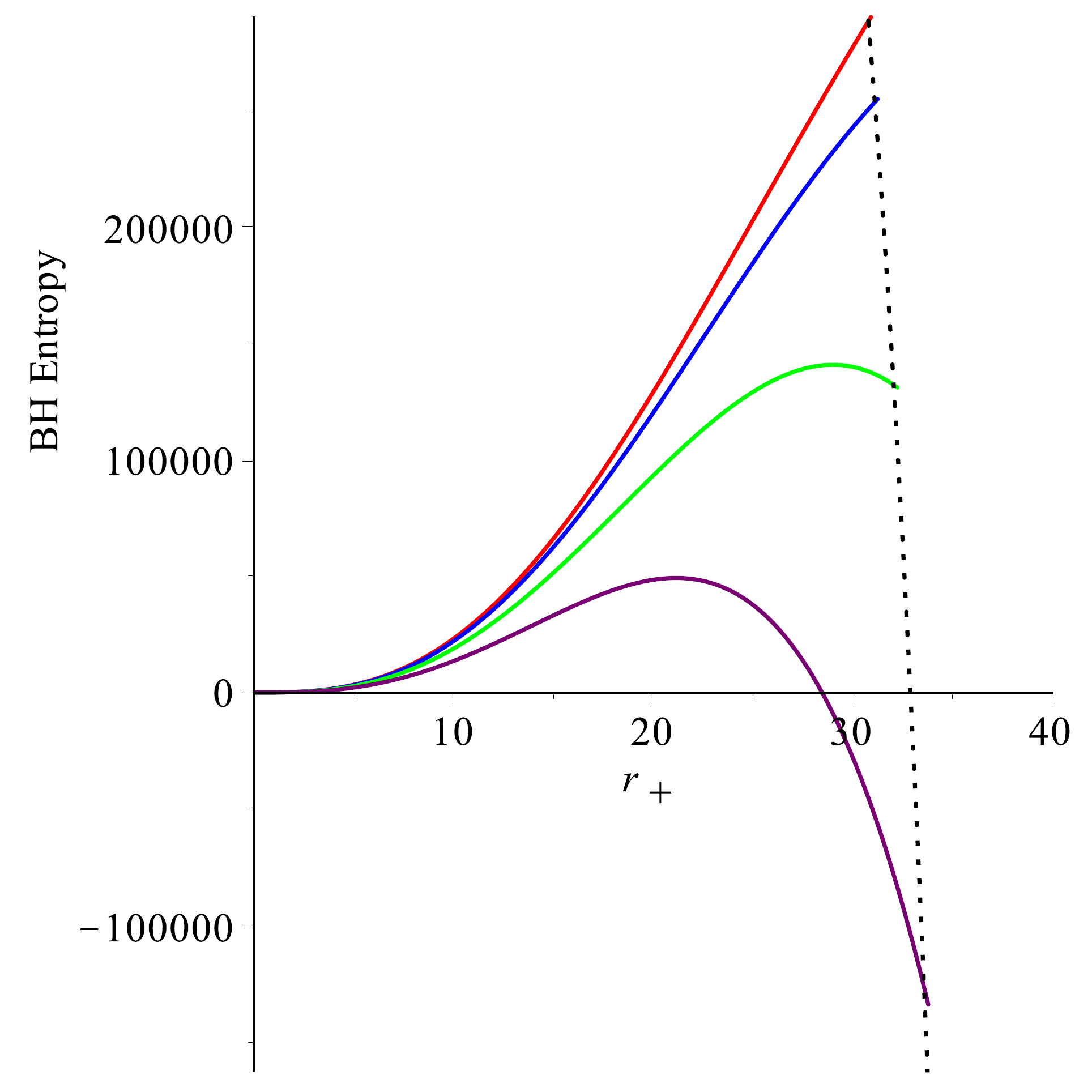}
\includegraphics[width=0.4\textwidth]{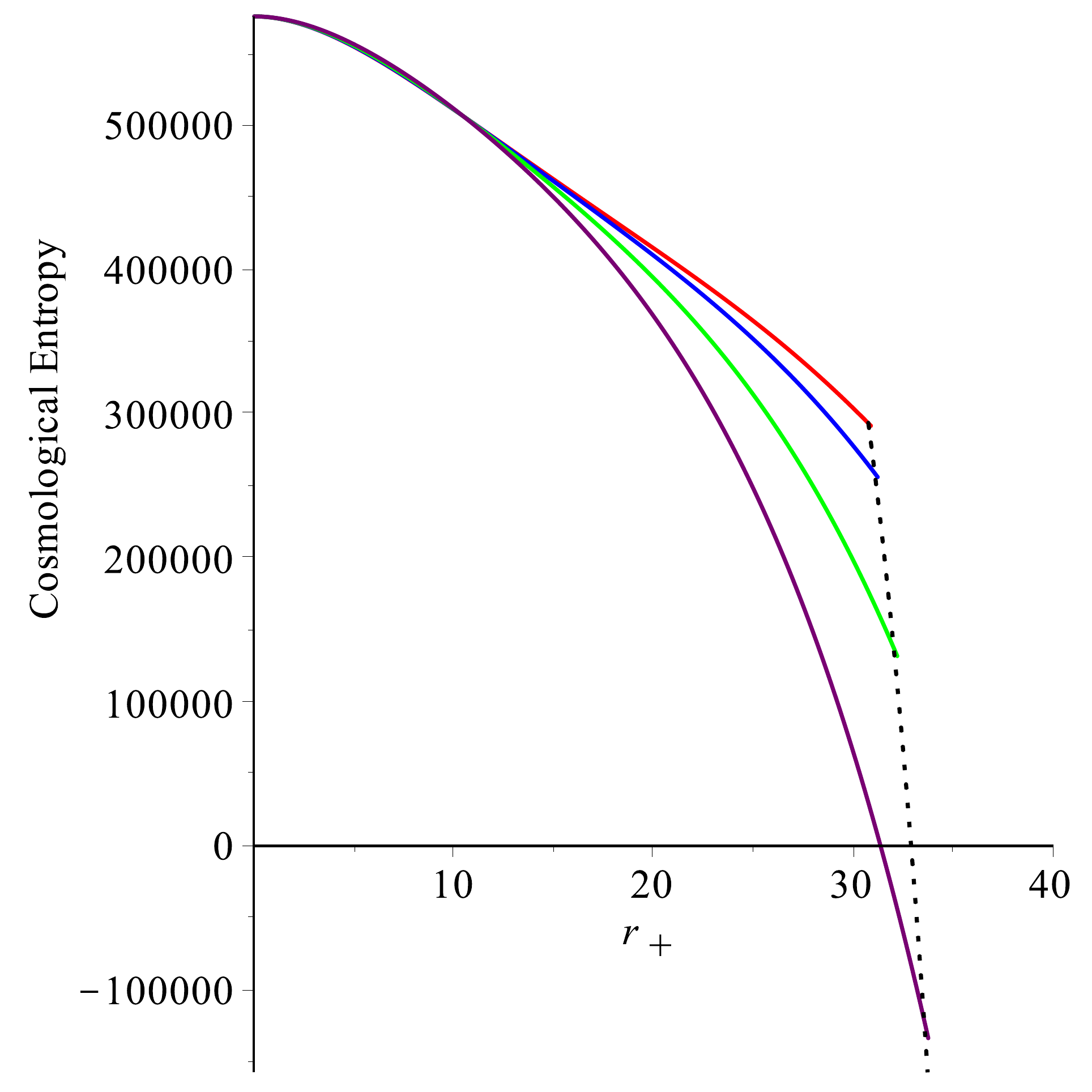}
\caption{The thermodynamic parameters of the charged hairy asymptotically de Sitter black holes: the case of $P= -0.0001$. The $ \{ red, blue, green, purple \} $ curves correspond to the potential $\phi = \{ 0.1 , 1 , 2 , 3 \}$ respectively . The dotted line represents the Nariai limit.
} 
\label{fig1}
\end{figure}
\begin{figure}[H]
\centering
\includegraphics[width=0.4\textwidth]{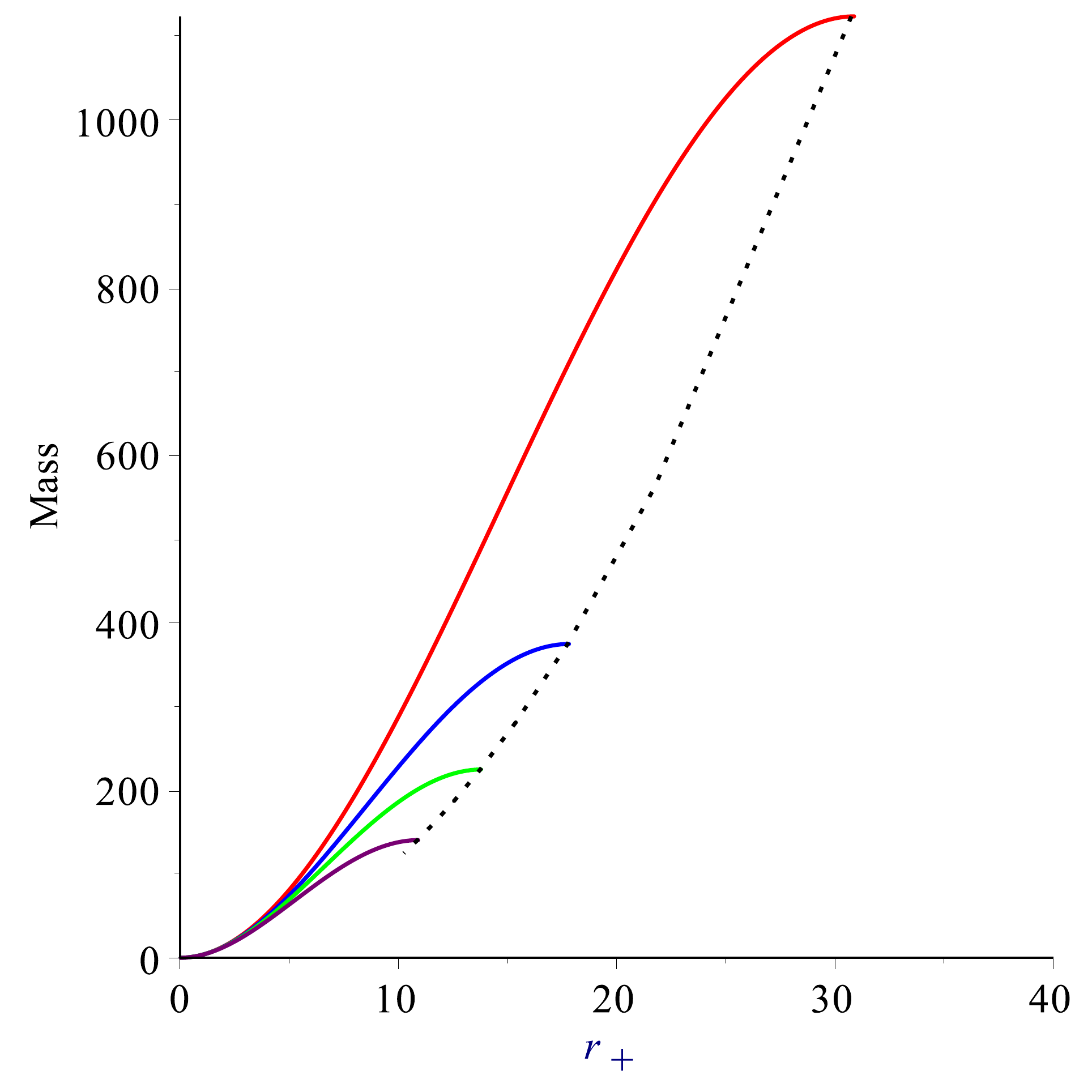}
\includegraphics[width=0.4\textwidth]{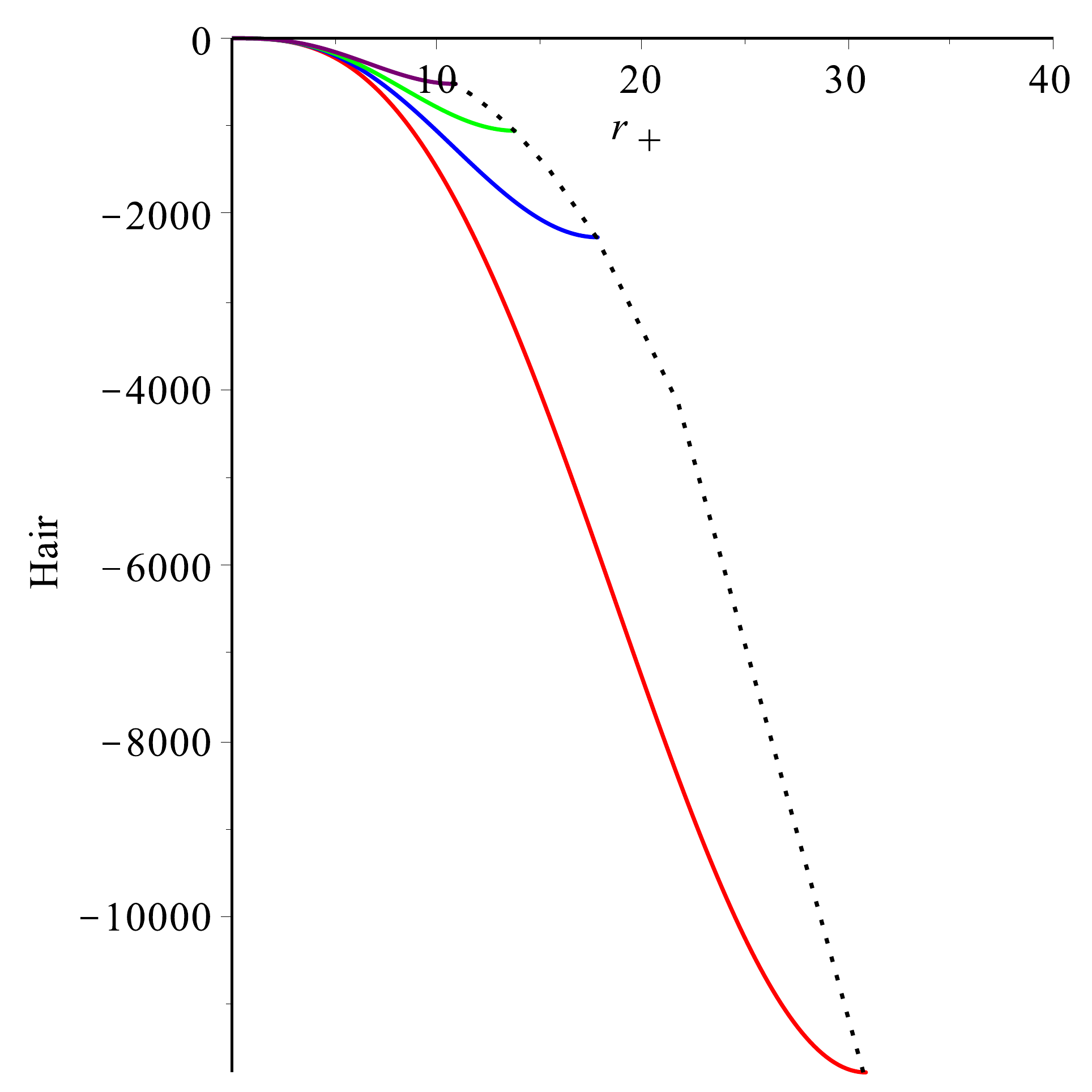}
\includegraphics[width=0.4\textwidth]{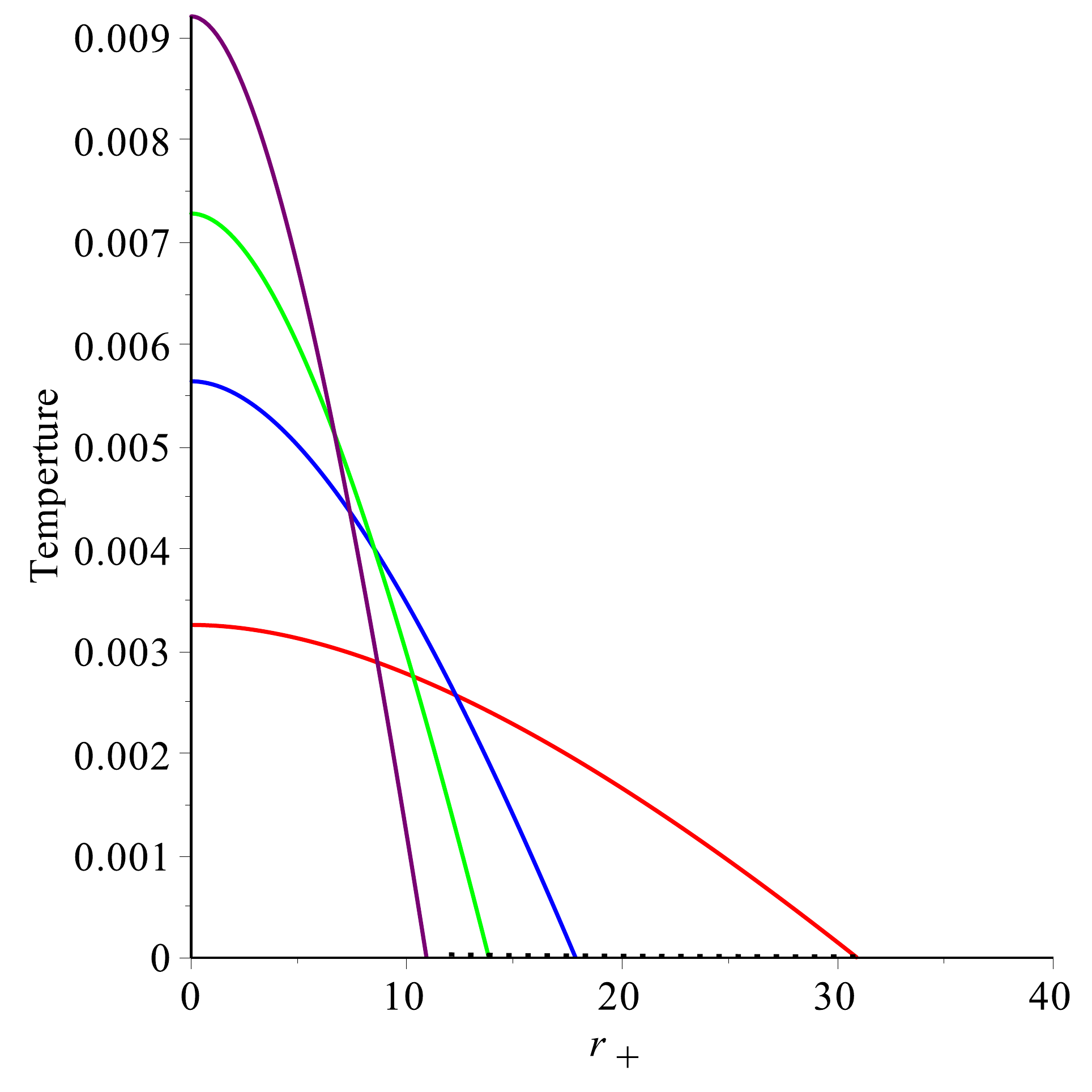}
\includegraphics[width=0.4\textwidth]{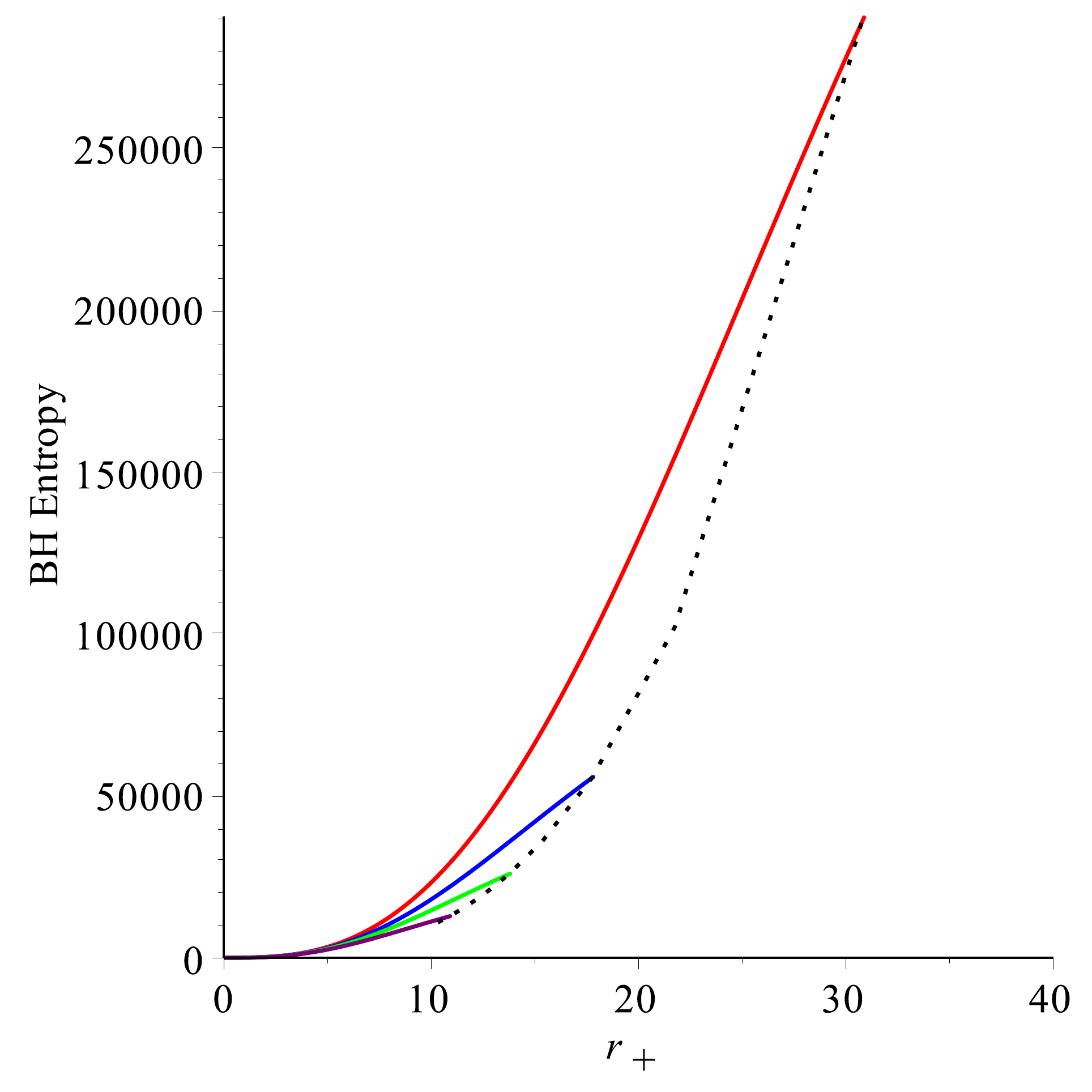}
\includegraphics[width=0.4\textwidth]{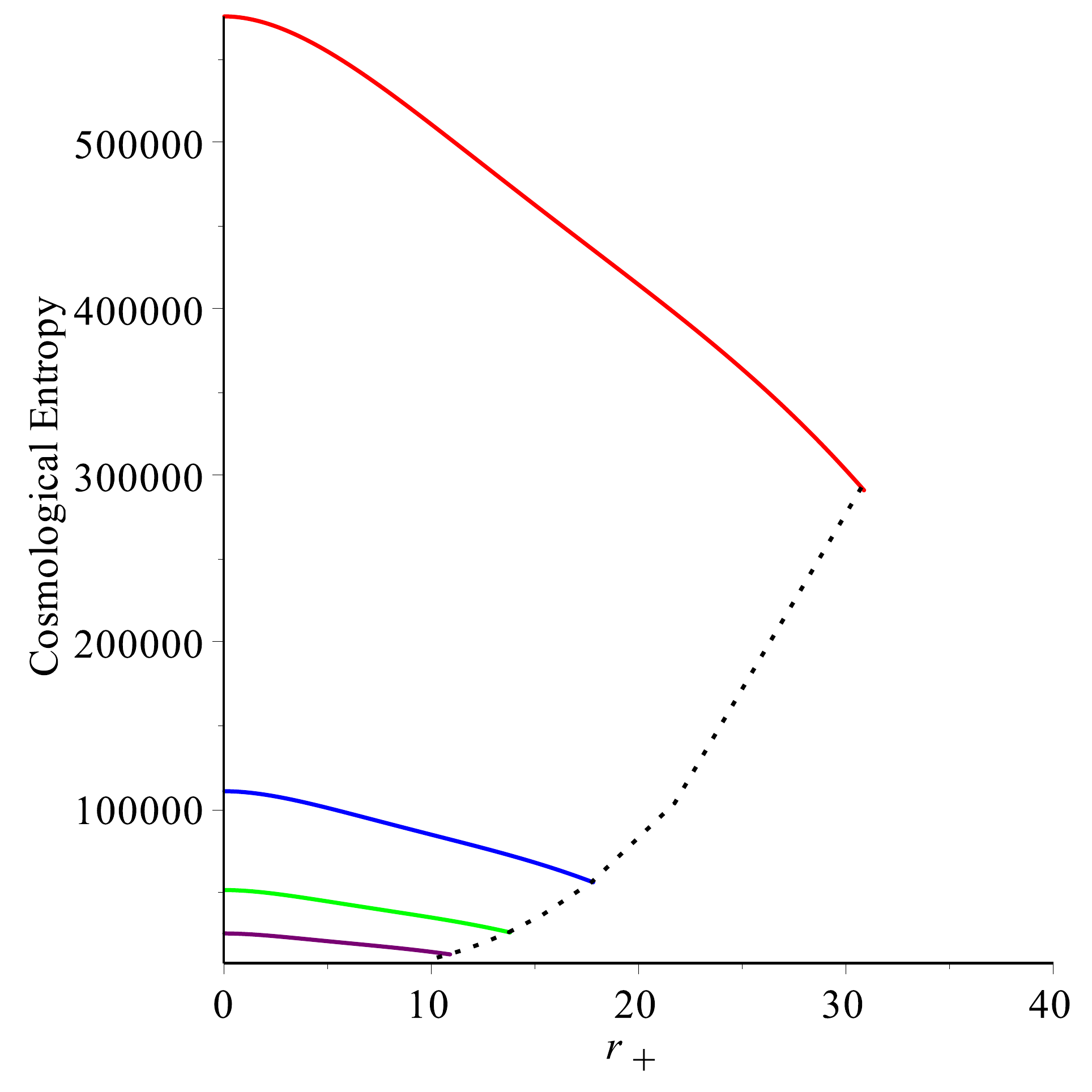}
\caption{The thermodynamic parameters of the charged hairy asymptotically de Sitter black holes for fixed potential  $\phi = -0.0001$. The $\{ red, blue, green, purple \}$ curves correspond respectively to the pressure  $ P = \{ -0.0001 , -0.0003 , -0.0005 , -0.0008 \}$. The dotted line represents the Nariai limit.
} 
\label{fig2}
\end{figure}

In what follows  we shall study  Asymptotically de Sitter charged hairy black holes in both the canonical and grand-canonical ensembles.

\subsection{Setup} 

 Naively, one would start the analysis of the chosen black holes at $d=4$. But since we are coupling the scalar field to the Gauss-Bonnet term, the latter makes no contribution to the field equations at $d=4$.  In other words, the metric function only receives a ``hairy" contribution when $d > 4$. All the expressions for the entropy in the previous section \eqref{entropy}  were derived assuming $d>4$.

We therefore specify to  $d=5$ Einstein gravity and we set $a_0 = -2 \Lambda < 0$, $a_1=1$,  $a_{k>1}=0$ and $G=1$ , obtaining  
\begin{equation}
f(r) = -\frac{1}{6} r^2 \Lambda +1 -\frac{H}{r^3} -\frac{8}{3}\frac{M}{\pi r^2}+\frac{4}{3} \frac{\pi Q^2}{r^4}
\label{metricF}
\end{equation}
 from \eqref{general_Poly}. At each horizon we have
\begin{equation}
-\frac{1}{6} \Lambda +\frac{1}{r_h^2} -\frac{H}{r_h^5} -\frac{8}{3}\frac{M}{\pi r_h^4}+\frac{4}{3} \frac{\pi Q^2}{r_h^6} = 0
\label{5d_Poly}
\end{equation}
and solve this polynomial equation, at both the event horizon $r_h=r_+$ and the cosmological horizon $r_h=r_c$, for the mass and the hair parameters. Both  can be regarded as functions of \{$r_c$, $r_+$, $Q$, $\Lambda$\} and so the  metric function $f$  now depends on these parameters. 
Since we are working in  extended phase space, we use the relationship between the cosmological constant and the pressure $\Lambda = -8\pi P$ (which, since $P<0$, is actually a tension) and then  solve \eqref{5d_Temp} for the cosmological horizon parameter $r_c$ where
\begin{align}
T_{+} &= - \frac{1}{12} \frac{ \Lambda  r{_+}  }{\pi} - \frac{4}{3} \frac{Q^2}{r_+^5} + \frac{3}{4} \frac{H}{\pi r_+^4} + \frac{4}{3} \frac{M}{\pi^2 r_+^3}  \nn \\
T_{c}  &= \frac{1}{12} \frac{ \Lambda  r{_c}  }{\pi} + \frac{4}{3} \frac{Q^2}{r_c^5} - \frac{3}{4} \frac{H}{\pi r_c^4} - \frac{4}{3} \frac{M}{\pi^2 r_c^3}  
\label{Temps}
\end{align}. 

The resultant equation has nine different roots but only one of them is a physical solution. The net result is that all thermodynamic parameters -- mass $M$, entropy $S$, temperature $T$,  and hair $H$ -- are now functions of  
\{$r_+$, $Q$, $P$\}.  This makes the analysis analogous to that of charged AdS black holes \cite{Kubiznak:2012wp}, and our setup can be understood and studied in several ways. The most relevant are understanding this system's thermodynamics in the contexts of the canonical ensemble and the grand-canonical ensemble.

\subsection{The Grand Canonical Ensemble: Thermodynamics with Fixed Potential}

The grand canonical ensemble is defined as the process that couples the energy and charge reservoirs of the system in question while holding the temperature and the potential fixed \citep{Chamblin:1999tk, Chamblin:1999hg}. Its corresponding thermodynamic potential is the known Gibbs free energy. 

The first question presented by our de Sitter black holes is that of which Gibbs free energy should we consider? 
There are three possibilities: that of the black hole, that of the cosmological horizon or that of the total system that includes the black hole and the cosmological horizon. These respectively read
\begin{align} \label{All-Gibbs} 
\ G_{+} &= M - T S_+ - \phi_+  Q   \nn \\
G_{c} &= M - T S_c - \phi_c  Q  \\
G_{Total} &= M - T (S_+ + S_c) - (\phi_+  - \phi_c) Q  \nn 
\end{align}
where $S_+$ is the entropy of the black hole and $S_c$ is the entropy at the cosmological horizon. Note that we regard the black hole as a thermodynamic system in equilibrium with the de Sitter vacuum. 
We plot $M$, $T$, and $S$  of the black hole as a function of $r_+$ in figures \ref{fig1}  and \ref{fig2}. We consider only those values of $r_+$ for which these quantities are positive, and ensure that  $r_+ < r_c$, so that we do not attain the Nariai limit.

\begin{figure}[h]
\centering
\includegraphics[width=0.4\textwidth]{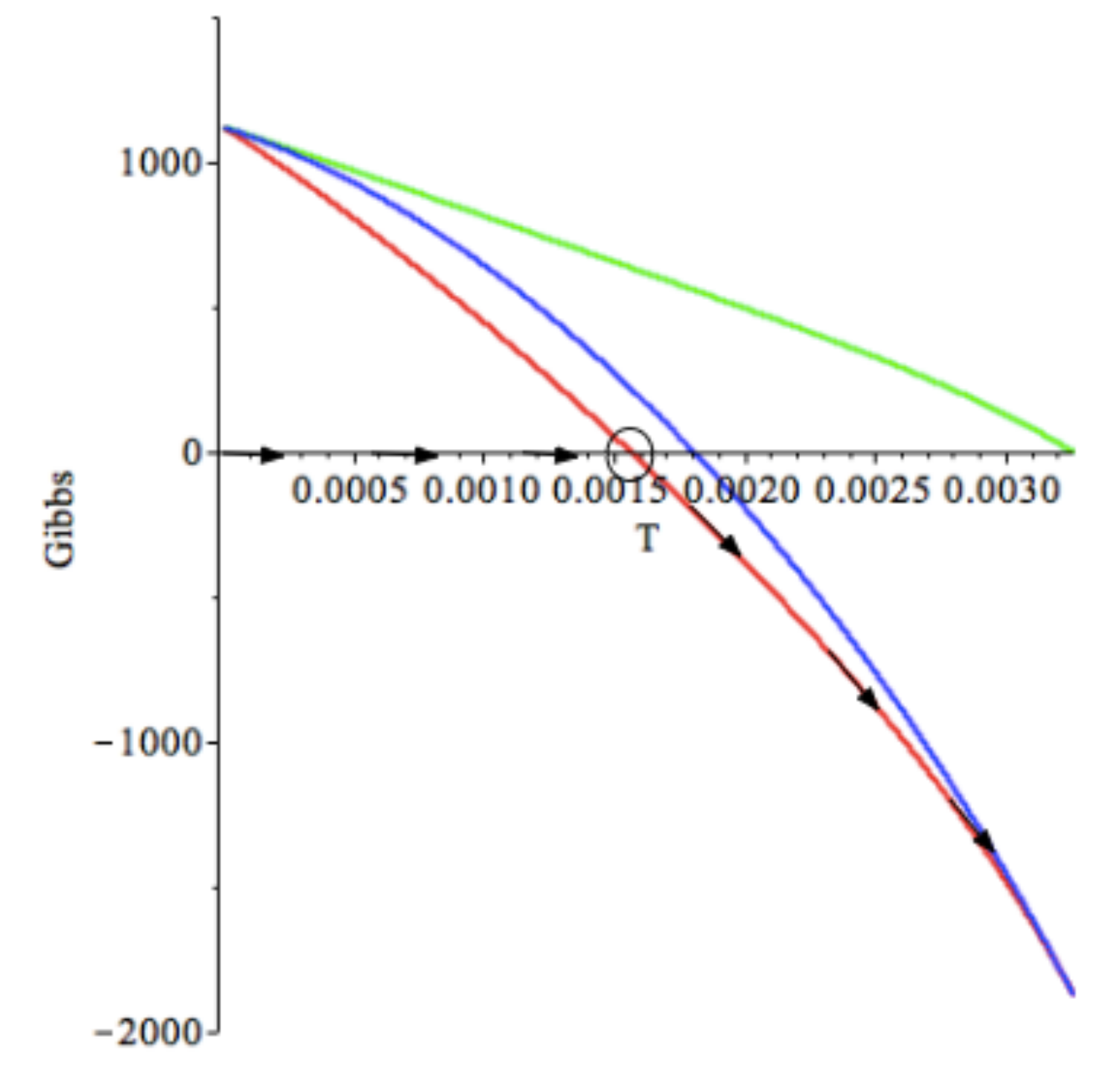}
\caption{{\bf Gibbs Free Energy with fixed potential}. The green curve represents   $G_+$, corresponding only to the black hole, as a function of temperature. The blue curve represents $G_c$ as measured at the cosmological horizon $r_c$ and the red curve is that of  $G_{Total} $ for the total system \textit{i.e.} of the black hole in a the de Sitter heat bath. All curves are plotted for $P= -0.0001$ and $\phi= 0.1$.
} 
\label{fig: Gibbs}
\end{figure}

Plotting the Gibbs free energy for the different cases in figure \ref{fig: Gibbs} we see that the black hole and the cosmological horizons cannot undergo any phase transitions if isolated from each other: the \textit{red} curve is the Gibbs free energy of the black hole when isolated from the cosmological horizon. On its own, it can't undergo a phase transition as the curve doesn't cross the plane of $G_{Total} = 0$. Nonetheless, the \textit{blue} curve that represents the free energy of the cosmological horizon does. Yet it can't undergo a phase transition since the cosmological  horizon can't be decoupled from the black hole (i.e. we use the mass of the black hole to compute the free energy of the cosmological horizon) and so has no physical meaning; the curve is plotted for reference.

\begin{figure}[h]
\centering
\includegraphics[width=0.4\textwidth]{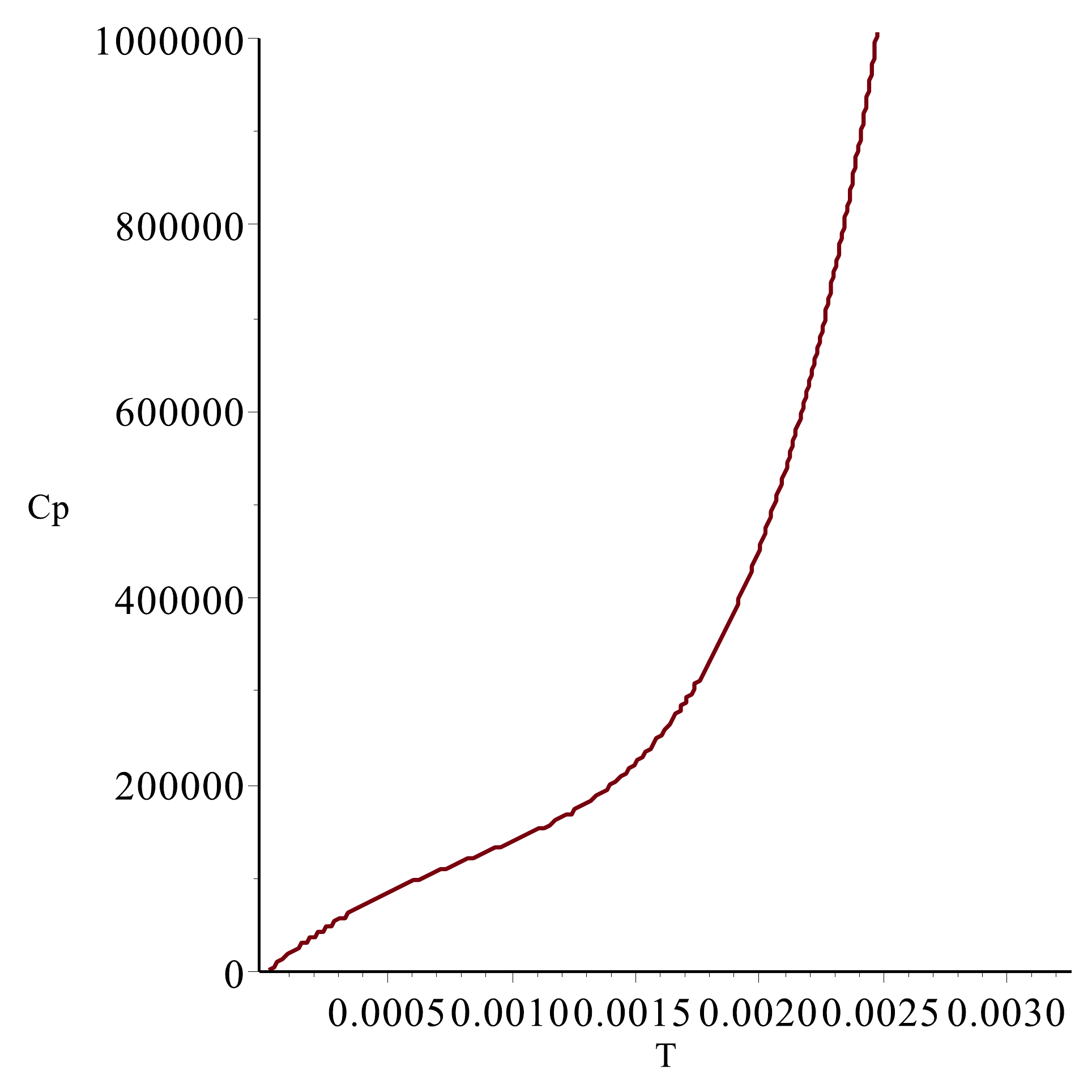}
\caption{{\bf Specific Heat}. This curve represents Cp, the specific heat of the system, as a function of the temperature. It is plotted for $P= -0.0001$ and $\phi= 0.1$.
} 
\label{fig: Cp}
\end{figure}

However,the full system undergoes a phase transition that resembles the Hawking-Page phase transition: at low temperatures, the equilibrium state of the system is de Sitter space with scalar radiation, and not a black hole. As the temperature increases, the system can undergo a first order phase transition to a new equilibrium state  of a black hole with scalar hair. The black hole would be of large size and would keep shrinking down as the system heats up. But at high temperatures, unlike the Schwarzschild black holes which have a negative specific heat and are unstable,  the charged and hairy   de Sitter black holes have a positive specific heat as shown in figure \ref{fig: Cp} and are stable -- as the small (high-T) black hole radiates, the cosmological horizon will restore the particle flux to ensure equilibrium.

The arrows in the plot \ref{fig: Gibbs} indicate  how the equilibrium state of the system changes 
from a state of scalar radiation in de Sitter spacetime to a  charged hairy black hole. Along the black hole
branch the arrows correspond to  decreasing values of the   radius of the black hole event horizon. This is quite
unlike the corresponding situation in anti de Sitter space \cite{Hawking:1982dh}, as well as for 
de Sitter black holes in cavities \cite{Carlip2003, Braden:1990hw,Simovic:2018tdy},
in which large black holes are at higher temperature. 
 We call this a \textit{Reverse } Hawking-Page phase transition.

\begin{figure}[h]
\centering
\includegraphics[width=0.4\textwidth]{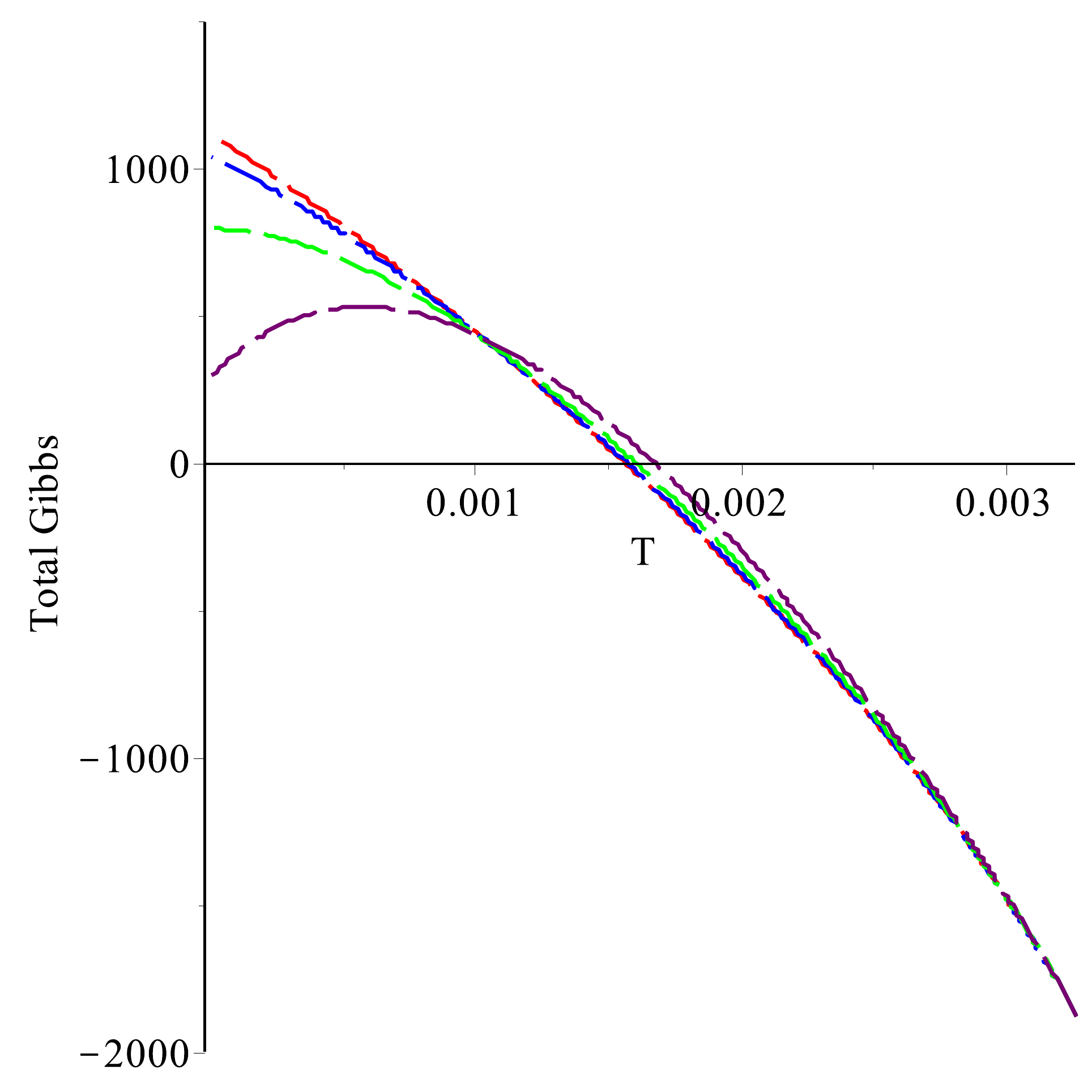}
\includegraphics[width=0.4\textwidth]{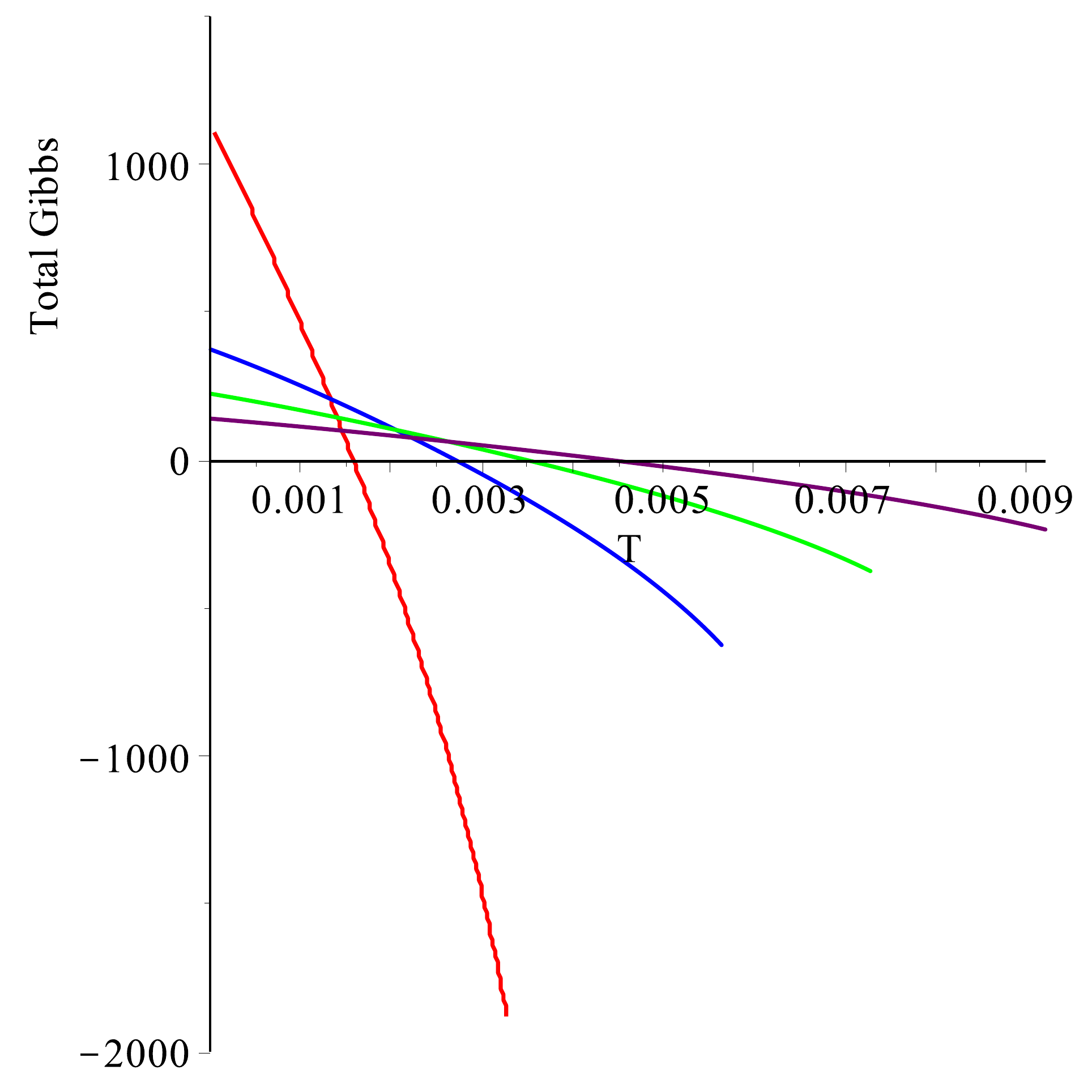}
\caption{{\bf Total Gibbs Free Energy $G_{Total} $}  \textit{Left:} $G_{Total} $ at $P = -0.0001$ where the $\{ red, blue, green, purple \}$ curves correspond to the potential $\phi = \{ 0.1 , 1 , 2 , 3 \}$ respectively . \textit{Right:} $G_{Total} $  at $\phi = 0.1$ where the $\{ red, blue, green, purple \}$ curves correspond to the pressure  $ P = \{ -0.0001 , -0.0003 , -0.0005 , -0.0008 \}$ respectively .
} 
\label{fig: Total_Gibbs}
\end{figure}

The behaviour of the total Gibbs free energy $G_{Total} $ for a variety of tensions and potentials, as seen in figure \ref{fig: Total_Gibbs}, shows that the system is sensitive to each. These two parameters have different effects as to the temperature at the phase transition can actually happen.  The transition temperature of the system is almost indifferent to the choice of fixed potential of the system but is highly sensitive to its tension. This is understandable as the effect of the charge located at the centre of the black hole shouldn't be significant, unlike the tension that is applied on the black hole by the de Sitter spacetime that effects the critical temperature dramatically.

 \subsection{Canonical Ensemble: Thermodynamics with Fixed Charge}
 
\begin{figure}[h]
\centering
\includegraphics[width=0.4\textwidth]{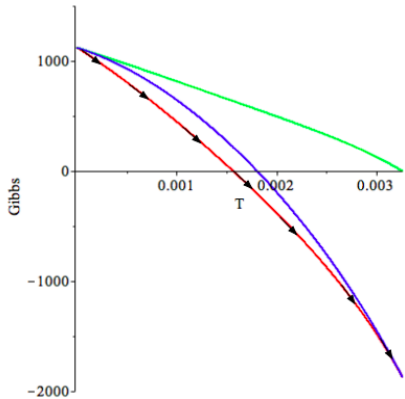}
\caption{  {Canonical Gibbs} Free Energy( {CGFE}): the green curve represents the  {CGFE} corresponding only to the black hole, the blue curve represents the  {CGFE} as measured at the cosmological horizon $r_c$, the red curve is of the  {CGFE} of the total system \textit{i.e.} of the black hole in a the de Sitter heat bath . All curves are plotted for $P= -0.0001$ and $Q= 1$.
} 
\label{fig: Canonical_Gibbs}
\end{figure}

The canonical ensemble, unlike the grand canonical ensemble, holds the charge fixed instead of the potential \citep{Chamblin:1999tk, Chamblin:1999hg}. However unlike these investigations, since 
the mass M as the enthalpy in the extended phase space, we consider minimization of the Gibbs free energy $F$ and not the Helmholtz free energy. Once again we consider the multiple scenarios: the free energy of the black hole, of the cosmological horizon, and  of the total system. They respectively read :
\begin{align} \label{All-Gibbs-Canon} 
\ F_{+} &= M - T S_+   \nn \\
F_{c} &= M - T S_+   \\
F_{Total} &= M - T (S_+ + S_c)  \nn 
\end{align}
As in  the previous section, we start by studying the different scenarios of different Gibbs free energies: The free energy of the black hole versus the free energy of the total system. We see in figure \ref{fig: Canonical_Gibbs} the behaviour of all three is respectively analogous to that found in  the grand canonical ensemble. However due to the conservation of charge, the total system will not undergo a phase transition at $F_{Total} = 0$ -- there will always be an equilibrium state consisting of a charged hairy black hole in de Sitter space. This system is always be stable, as its specific heat is always positive (the plot of the specific heat for this case is similar to that in figure \ref{fig: Cp}).

The behaviour of the total  {Gibbs} free energy $F_{Total} $ for a variety of tensions and charges showed similar results to \ref{fig: Total_Gibbs}. These two parameters have different effects on the free energy.  But as there's no phase transition possible this sensitivity is of marginal interest.

\section{Conclusion} 

We have for the first time studied the thermodynamics of a black hole with two horizons with thermodynamic equilibrium imposed and with a control parameter.  This study is possible because this class of black holes has an additional degree of freedom given by a hair parameter that allows us to impose thermodynamic equilibrium. 

In addition of thoroughly studying the thermodynamic parameters of these black holes in de Sitter, we have analyzed the free energy of these systems in different ensembles. We have found that the system can undergo a phase transition that resembles the Hawking-Page phase transition in the grand-canonical ensemble, a result that is consistent with results previously found for de Sitter black holes  isolated  inside a cavity with thermodynamic equilibrium externally imposed \cite{Carlip2003}.  However a key difference in our result is that   at high-T  the small black hole remains stable as it radiates,  the cosmological horizon restoring the particle flux; for this reason we  call it a ``Reverse Hawking-Page'' phase transition. We also found that these systems can not undergo any phase transition in the canonical ensemble as such a transition violates the conservation of charge.

The situation here also stands in notable contrast to a recent study of the behaviour of charged de Sitter black holes in a cavity that takes into account pressure (tension) and volume \cite{Simovic:2018tdy}.  In this case the cavity is used to ensure equilibrium, and one finds not only a standard Hawking Page transition but also 
a Van der Waals transition that exists only for a finite range of non-zero pressure, described by a ``swallowtube" 
structure in a plot of free energy vs. pressure and temperature.   However we find that this structure does not appear when   scalar hair is used to 
ensure equilibrium, and only a reverse Hawking-Page transition is
possible.

Ours is an exceptional setup that allows us to uncover the thermodynamic properties and phase transitions that call tell us more about de Sitter space.    It is certainly very important to further study these classes of black holes  in higher dimensions and in other higher-curvature theories of gravity to see what other interesting phase behaviour might be present.

\section*{Acknowledgments}
This work was supported in part by the Natural Sciences and Engineering Research Council of Canada.   We are grateful to B.Y. Ha , R. A. Hennigar and D. Kubiznak for helpful discussions. 
 
\bibliographystyle{JHEP}
\bibliography{LBIB2}  
\end{document}